\documentclass[aps,prc,longbibliography,reprint,superscriptaddress,twoside,twocolumn,nofootinbib,showpacs]{revtex4-1}

\usepackage{graphics}
\usepackage{graphicx}
\usepackage{epsfig}

\usepackage{color}
\usepackage{slashed}
\usepackage{multirow}
\usepackage{mathtools, mathrsfs, braket, tensor}
\usepackage{bm}
\usepackage{slashed}
\usepackage{booktabs}
\usepackage{color}

\newcommand{\tcb}{\textcolor{black}}

\usepackage{amssymb, amsmath, latexsym}
\usepackage{eqnarray}

\newcommand{\EPA}{\mathcal{E}}   
\newcommand{\KEA}{\mathcal{T}}   

\begin{document}

\title{Constraining the density dependence of the symmetry energy with nuclear data and astronomical observations  in the Korea-IBS-Daegu-SKKU framework} 


\author{Hana Gil}
\email{khn1219@gmail.com}
\affiliation{Center for Extreme Nuclear Matters, Korea University, Seoul 02841, Korea}
\author{Young-Min Kim}
\email{ymkim715@gmail.com}
\affiliation{Department of Physics, Ulsan National Institute of Science and Technology, Ulsan 44919, Korea}
\author{Panagiota Papakonstantinou}
\email{ppapakon@ibs.re.kr}
\affiliation{Rare Isotope Science Project, Institute for Basic Sceince,
Daejeon 34047, Korea}
\author{Chang Ho Hyun}
\email{hch@daegu.ac.kr}
\affiliation{Department of Physics Education,
Daegu University, Gyeongsan 38453, Korea}

\date{\today}

\begin{abstract}
\noindent
{\bf Background:} The properties of very neutron rich nuclear systems are largely determined by the density dependence of the nuclear symmetry energy. 
The Korea-IBS-Daegu-SKKU (KIDS) framework for the nuclear equation of state (EoS) and energy density functional (EDF) offers the possibility to explore the symmetry-energy parameters such as $J$ (value at saturation density), $L$ (slope at saturation), $K_{\rm sym}$ (curvature at saturation) and higher-order ones independently of each other and independently of assumptions about the in-medium effective mass, as 
previously shown in the cases of closed-shell nuclei and neutron-star properties. \\
{\bf Purpose:} We examine the performance of  EoSs with different symmetry energy parameters on properties of nuclei and observations of neutron stars and gravitational waves  in an effort to constrain in particular $L$ and $K_{\rm sym}$ or the droplet-model counterpart $K_{\tau}$.   
\\
{\bf Method:} 
Assuming a standard EoS for symmetric nuclear matter, we explore several points on the hyperplane of $(J,L,K_{\rm sym}\, {\rm or} \, K_{\tau})$ values. 
For each point, the corresponding KIDS functional parameters and a pairing parameter are obtained for applications in spherical even-even nuclei. 
This is the first application of KIDS energy density functionals with pairing correlations in a spherical Hartree-Fock-Bogoliubov (HFB) computational code.
The different EoSs are tested successively on the properties of closed-shell nuclei, along the Sn isotopic chain, and on astronomical observations, in a step-by-step procedure of elimination and correction. \\ 
{\bf Results:}  
A small regime of best-performing  parameters is determined and correlations between symmetry-energy parameters are critically discussed. 
The results strongly suggest that $K_{\rm sym}$ is negative and no lower than $-200$~MeV, that $K_{\tau}$ lies between roughly $-400$ and $-300$~MeV 
and that $L$ lies between 40 and 65~MeV, with $L\lessapprox 55$MeV more likely. 
For the selected well-performing sets,  corresponding predictions for the position of the neutron drip line and the neutron skin thickness of selected nuclei are reported. 
The results are only weakly affected by the choice of effective mass values. Parts of the drip line can be sensitive to the symmetry energy parameters.
\\
{\bf Conclusion:} 
Using KIDS EoSs for unpolarized homogeneous matter at zero temperature and KIDS EDFs with pairing correlations in spherical symmetry 
we have explored the hyperplane of symmetry-energy parameters. 
Using both nuclear-structure data and astronomical observations as a testing ground, 
a narrow regime of well-performing parameters has been determined, free of non-physical correlations and decoupled from constraints on the nucleon effective mass. 
The results underscore the role of $K_{\mathrm{\tau}}$ and of precise astronomical observations. 
More-precise constraints are possible with precise fits to nuclear energies and, in the future, more-precise input from astronomical observations. 
\end{abstract}
\maketitle

\section{Introduction}
\label{sec:intro}

The number of measured or observed nuclei exceeds 3,000, providing firm ground to test
models and theories of nuclear structure.
At the same time, rare-isotope accelerators in operation or under construction around the world
are expected to explore the nuclear landscape even further towards the drip lines, 
while astronomical observations including measurements of gravitational waves usher in a new epoch for the understanding of neutron stars.
Such progress demands a more precise description of neutron-rich and/or high-density nuclear systems and a better understanding of the uncertainties and biases involved in the models. 
A key issue is the the density dependence of the nuclear symmetry energy~\cite{review2016,Roca2018}, 
typically represented by the parameters characterizing its value and derivatives at the saturation point, 
\begin{eqnarray}
S(\rho) = J + L x + \frac{1}{2} K_{\rm sym} x^2 + \frac16 Q_{\rm sym} x^3 + \cdots,
\label{eq:symene}
\end{eqnarray}
where $S$ denotes the symmetry energy, $\rho$ is the nucleon density, and $x = (\rho - \rho_0)/3 \rho_0$ 
with $\rho_0$ the nuclear saturation density.
Representative ways or observables to constrain the parameters include the neutron skin thickness of neutron-rich nuclei,
isobaric analog states, heavy ion collisions, and information from astrophysics~\cite{Lattimer2013,review2016,Oertel2017,Roca2018,Agrawal2020p}. 
Uncertainties are large especially for the higher-order parameters.  
There are some overlapping acceptable ranges for $J$ and $L$ coming from diverse sources. 
For example, from the results reviewed in Ref.~\cite{review2016}, one can infer the ranges $32~{\rm MeV} < J <33$~MeV, and $55~{\rm MeV} < L < 60$~MeV \cite{review2016}. 
On the other hand, an averaging performed in Ref.~\cite{Oertel2017} of about 50 constraints reported in the literature resulted in the ranges $J=31.7\pm3.2$~MeV and $L=58.7\pm 28.1$~MeV. 
Higher-order parameters such as $K_{\rm sym}$ remain practically unconstrained. 
Efforts to constrain their values largely rely on correlations observed between $K_{\rm sym}$ and $J$, $L$ values in model calculations. 
Correlations are mostly provided by traditional energy-density functional (EDF) models and it is unclear if they are model-specific. 
Therefore it is worthwhile to check the ranges in terms of new ways, in particular explore the symmetry-energy parameters independently of each other. 
The so-called meta-model proposed in Ref.~\cite{Mar2018} for the equation of state (EoS) aims to serve such a purpose and has found several interesting applications in studies of homogeneous matter and in analyses of astronomical observations.  On the other hand, \tcb{the meta-model} is not easy to apply to nuclei beyond the Thomas-Fermi approximation~\cite{meta_nuclei1}, 
\tcb{which may not be appropriate for an accurate description of finite nuclei. In} addition, there are diverse strategies for correcting the \tcb{meta-model} EoS near zero density
 - see also Ref.~\cite{Tsang2020} about this issue. 

The Korea-IBS-Daegu-SKKU (KIDS) framework is based on a natural ansatz for the EoS~\cite{kidsnm} and provides a clear strategy for transposing any EoS into a \tcb{microscopic} EDF for finite nuclei~\cite{kids_nuclei1,kids_nuclei2,Gil2017}. 
In addition, assumptions for the EoS parameters are independent of assumptions for the in-medium effective mass. 
By EoS we mean here the energy per particle of homogeneous matter at zero temperature, while EDF is a corresponding functional model for applications in nuclear structure. 
The KIDS framework allows us to test EoS parameters on both infinite matter and finite nuclei and was developed with such explorations  in mind.
In particular, any given EoS can be tested on basic nuclear observables and judged realistic or not according to its performance.

In this work, we examine the performance of  EoSs with different symmetry energy parameters on properties of nuclei and observations of neutron stars and gravitational waves  in an effort to constrain in particular $L$ and $K_{\rm sym}$ or the droplet-model counterpart $K_{\tau}$. 
Assuming a standard EoS for symmetric nuclear matter, we explore several points on the hyperplane of $(J,L,K_{\rm sym}\, {\rm or} \, K_{\tau})$ values. 
For each point, the corresponding KIDS functional parameters and a pairing parameter are obtained for applications in spherical even-even nuclei. 
This is the first time that KIDS energy density functionals are applied in spherical Hartree-Fock-Bogoliubov (HFB) calculations taking into account pairing correlations. 
The different EoSs are tested successively on basic properties of closed-shell nuclei, along the Sn isotopic chain, and on astronomical observations,
 in a step-by-step procedure of elimination and correction. 
Finite nuclei represent the density regime up to the saturation point, while astronomical observations provide constraints for the higher-density regime. 
A restricted domain of best-performing  parameters is determined and correlations between symmetry-energy parameters are discussed. 
As we will see, 
traditional  EDF models, such as Skyrme models with only one density-dependent term, are unsuitable for examining the role of all three parameters, because they force an almost analytical relation between the three EoS parameters owing to the insufficient number of corresponding EDF parameters. 
We believe this deficiency to belie paradoxical findings such that ``realistic" EoS parameters cannot necessarily be applied in the description of nuclei with realistic results (see, e.g., \cite{Roca2018}).
For the selected well-performing sets, we also report predictions for the position of the neutron drip line and the neutron skin thickness of selected nuclei. 

The paper is organized as follows. 
We begin in Sec.~\ref{Sec:KIDS} with a reminder of the basic principles of the KIDS framework, now extended to include pairing, and some comments on correlations among  EoS parameters in traditional Skyrme models. 
In Sec.~\ref{Sec:Method} we undertake an exploration of the $(J,L,K_{\tau})$ hyperplane centered around values of $(J,L)$ generally considered reasonable. 
We test the EoSs successively on  basic properties of closed-shell nuclei, of open-shell nuclei and of neutron stars, in a step-by-step process of elimination and correction. 
In  Sec.~\ref{Sec:Results} we compile our results, revisit the aforementioned correlations between EoS parameters and provide predictions for the neutron drip line and the neutron-skin thickness. 
We summarize our work and discuss future perspectives in Sec.~\ref{Sec:End}.

\section{From EoS to EDF and nuclei\label{Sec:KIDS}}

First, we summarize the basic features of the KIDS EoS and the KIDS EDF. 
Next, we introduce the application to a spherical HFB code. 
We will also comment on a correlation which has been observed between symmetry energy parameters in traditional EDF models and which we will revisit in this work. 

\subsection{KIDS framework \label{Sec:KIDSfm}}
In infinite nuclear matter the energy per particle $\EPA$ at given isospin asymmetry $\delta= (\rho_n - \rho_p)/\rho$ is expanded in powers of the cubic root of the density $\rho=\rho_p+\rho_n$, with $\rho_n$ and $\rho_p$ the neutron and proton density, respectively. 
The form and most relevant terms were determined and discussed in Ref.~\cite{kidsnm}. 
If we assume in addition a quadratic dependence of the potential energy on $\delta$, the energy per particle of infinite matter is written as
\begin{eqnarray}
{\EPA}(\rho,\, \delta) = {\cal T}(\rho,\, \delta) + \sum^2_{i=0} \alpha_i \rho^{1+i/3} 
+ \delta^2 \sum_{i=0}^3 \beta_i \rho^{1+i/3},
\label{Eq:EoS} 
\end{eqnarray}
where ${\KEA}(\rho,\, \delta)\propto \rho^{2/3}$ 
is the kinetic energy per particle of a free Fermi gas. 
The optimal number of terms was found to be three terms in the isospin symmetric part, $\alpha_i$, 
and 4 terms in the isospin asymmetric part, $\beta_i$~\cite{kidsnm,kids_nuclei2}.

With three parameters in the symmetric part $\alpha_i$, three EoS parameters of symmetric nuclear matter (SNM), typically the saturation density $\rho_0$, the energy per particle at saturation $\EPA_0$, and the compression modulus $K_0$, can be controlled independently. 
The relation between $(\rho_0,\EPA_0,K_0)$ and $(\alpha_0,\alpha_1,\alpha_2)$ is analytical and straightforward; 
similarly, with four parameters $\beta_i$ four symmetry-energy parameters can be controlled, i.e., of $J$, $L$, $K_{\rm sym}$, and $Q_{\rm sym}$ in Eq. (\ref{eq:symene})  \cite{kids_nuclei2}.
The analytical relations between the KIDS-EoS parameters ($\alpha_i, \beta_i$) and the characteristic EoS parameters  $(\rho_0,\EPA_0,K_0,J,L$, etc.) can be found in 
Refs.~\cite{kids_nuclei2,PP2018HNPS}.

Given an EoS $(\rho_0,\EPA_0,K_0,J,L,K_{\rm sym},Q_{\rm sym})$ under investigation, a corresponding KIDS EDF can be obtained as explained in previous work for applications in nuclei~\cite{kids_nuclei1,kids_nuclei2,PP2018HNPS}. 
The KIDS EDF is characterized by Skyrme-like parameters 
$(t_i, y_i, W_0)$, 
where $i$ stands for the indices $(0,1,2,31,32,33)$, $y_i$ replaces the Skyrme notation $t_ix_i$ for the exchange terms, and $W_0$ is a spin-orbit parameter. 
The indices $31,32,33$ represent three density-dependent couplings. 
All KIDS-EDF parameters except $W_0$ and four linear combinations of $t_1,y_1,t_2,y_2,t_{32},y_{32}$ can be determined analytically from the KIDS-EoS parameters.  
The analytical expressions connecting the KIDS-EDF parameters with the KIDS-EoS parameters ($\alpha_i,\beta_i$) can be found in Refs.~\cite{kids_nuclei1,PP2018HNPS}. 
Specific values for the 
isoscalar and isovector effective mass, $\mu_s=m^{\ast}/m$, $\mu_v=m_{\rm IV}^{\ast}/m$, may be specified at this stage, if desired, reducing the unknown linear combinations 
of \tcb{$t_i,y_i$} parameters to two.
\tcb{Specifically, the $\mu_s$ and $\mu_v$ values uniquely determine the combinations }
\[ 
\theta_s \equiv  3t_1+5t_2+4y_2,\quad 
\theta_{\mu} \equiv  t_1+3t_2-y_1+3y_2 
\] 
\tcb{and, with $\alpha_2,\beta_2$ known, they also determine $t_{32},y_{32}$~\cite{kids_nuclei1}. 
The two linear combinations related to the gradient terms of the functional, namely } 
\[
C_0^{\Delta\rho} \equiv \frac{1}{64} [-9t_1+5t_2+4y_2] , C_{1}^{\Delta\rho} \equiv \frac{1}{64} [3t_1+6y_1+t_2+2y_2]  
,\] 
\tcb{remain to be fitted.
(The above expressions for the effective mass terms and for the gradient coupling constants are exactly as encountered in the well-known Skyrme-functionals -- see, e.g., Appendix A of the review \cite{Bender2003}).  
Thus there remain three parameters in total to be determined from nuclear data, namely $C_0^{\Delta\rho}, C_1^{\Delta\rho}, W_0$. 
We note that it is straightforward to determine $t_{1,2}$ and $y_{1,2}$ from their independent linear combinations $\theta_s, \theta_\mu , C_0^{\Delta\rho}, C_1^{\Delta\rho}$.  
} 

Having found that the effective mass plays a minor role in predictions for bulk static properties (total energy and radius) of closed-shell nuclei, one can set $y_1=y_2=0$ and use the simpler method of Ref.~\cite{kids_nuclei2,Gil2017} in such cases.   
On the other hand, when spectroscopic precision is required, as in the case of (two-) nucleon separation energies, 
assumptions about the effective mass can be of importance. We will return to this point.

For determining the remaining KIDS-EDF parameters which are not determined from the EoS, input from nuclear data is needed. 
In this work, the input data are the binding energies and charge radii of closed-shell nuclei and the cost function is 
\begin{equation} 
{\psi^2}=\frac{1}{N}\sum_{n=1}^N\left( \frac{O^{\mathrm{calc}}_n-O^{\mathrm{exp}}_n}{O^{\mathrm{exp}}_n}  \right)^2
\label{Eq:ChiSq}
\end{equation}
with $N$ the number of input data. 
In the above, $O$ represents the energy or charge radius and the superscript the calculated or experimental value.

The stand-out feature of the above fitting procedure is that the EoS parameters are not adjusted in the process but remain immutable. 
The purpose of the fitting is to optimize the gradient and spin-orbit terms, which are not active in homogeneous matter. 
Another important feature of the KIDS framework is that assumptions for the EoS parameters are independent of assumptions for the in-medium effective mass. 
\tcb{
Unlike a standard Skyrme functional, here, any variations in $t_1, t_2$ and $y_1, y_2$ owing to the choice of effective mass are offset by a corresponding variation 
of $t_{32}$ and $y_{32}$ in the density dependence, so that the total homogeneous-matter EoS terms proportional to $\rho^{5/3}$ ($i=2$ in in Eq. (\ref{Eq:EoS})) remain the same. 
Thus in the KIDS framework, the effective mass is decoupled from the EoS of {\em homogeneous} matter and therefore cannot contaminate our constraints on the latter  -- as could be the case in more restricted 
ansaetze~\cite{Tsang2019,Zhang2020}, see also Sec.~\ref{Sec:Correl}. Interestingly, as already mentioned, the effective mass is also practically decoupled from bulk, static properties of magic nuclei [10]. On the other hand, it may affect predictions for open-shell nuclei (e.g., by influencing the single-particle spectrum and thus the pairing energy). Therefore, we will examine its effect again here, see Sec.~\ref{Sec:OSN} and Sec.~\ref{Sec:Drip}.  
} 

Once the KIDS EDF has been determined as above, its predictive power can be tested on more data. 
As a measure of predictive power we use the average deviation per datum (ADPD), 
\begin{equation} 
\mathrm{ADPD} = \frac{1}{N} \sum_{n=1}^{N} \left| \frac{O_n^{\mathrm{calc}} - O_{n}^{\mathrm{exp}}}{O_{n}^{\mathrm{exp}}} \right|  
\label{Eq:ADPD} .
\end{equation} 
In principle, the ${\psi^2}$ calculated on the new data set could also be used to the same effect. 

These are the basic elements of the KIDS EoS and EDF formalism as has been applied in closed-shell nuclei as well as neutron stars~\cite{kidsnm,kids_nuclei1,kids_nuclei2}. 
We should note that in previous work $N=6$ input data were used in Eq.~(\ref{Eq:ChiSq}) for deteremining an EDF from an EoS. 
The data were the binding energies and charge radii of $^{40,48}$Ca and $^{208}$Pb. 
Next the the predictive power of each parameterization, 
which is interpreted as how realistic the starting EoS is, could be tested on basic properties of several other nuclei. 
Here we will use the same strategy in an initial broad-range search for well-performing regions of the symmetry energy parameter space. 
Specifically, in Sec.~\ref{Sec:coarse} we will 
perform fits (${\psi^2}$, Eq.~(\ref{Eq:ChiSq})) on the above 6 data and check the performance (ADPD, Eq.~(\ref{Eq:ADPD})) on additionally 7 data ($N=13$ in total). 
The additional data are the binding energies and radii of $^{16}$O, $^{90}$Zr and $^{132}$Sn and the binding energy of $^{218}$U. 
Next, in Sec.~\ref{Sec:fine}, we will refine the search by directly fitting (${\psi^2}$, Eq.~(\ref{Eq:ChiSq})) to $N=13$ data. 
The performance of selected parameter sets (ADPD, Eq.~(\ref{Eq:ADPD})) will be examined in Sec.~\ref{Sec:Sn} on the energies of Sn isotopes ($N=20$ data). 
Finally we will examine neutron-star properties (Sec.~\ref{Sec:Results}). 

\tcb{The above choices for the cost function ${\psi^2}$ and ADPD, based on relative deviations,introduce some bias towards reproducing the total energy and the radius of lighter nuclei.
 Lighter nuclei can be useful for determining the gradient terms. Arguably, $^{16}$O in particular might be too light to be accurately described by the KIDS functional. However, the effect of including in our data set the isospin-symmetric nucleus $^{16}$O on results for the symmetry energy and neutron-rich nuclei is expected to be marginal. 
} 

So far the KIDS framework has been applied only to closed-shell nuclei. 
We proceed to explain how the KIDS EDF can be applied to spherical, even-even open shell nuclei for the purposes of this work.

\subsection{Open-neutron-shell even-even nuclei \label{Sec:OSN}}

Thus far the KIDS formalism has been applied in closed shell nuclei. 
It is rather straightforward to extend the applications to spherical, even-even open-shell nuclei. 
The KIDS EDF has the form of an extended Skyrme functional, i.e., a Skyrme functional with more than one density-dependent terms. 
Therefore, any computational code available for Skyrme functionals can be extended for KIDS applications quite easily. 
For spherical, even-even open-shell nuclei we have adopted the publically available HFB code HFBRAD (v1.0)~\cite{hfbrad} and extended it for use with KIDS functionals. 
Pairing correlations are accounted for by means of a two-nucleon pairing potential~\cite{hfbrad} 
\begin{eqnarray}
V_{\rm pair} = t_0'\left( 
 1 - \frac{\rho}{2\rho_0}  \right) 
\delta (\vec{r}_1 - \vec{r}_2).
\label{Eq:Vpair}
\end{eqnarray} 
In the notation of Ref.~\cite{hfbrad} and for $\rho_0=0.16~$fm$^{-3}$ the above corresponds to $\gamma'=1$ and $t_3'=-37.5t_0'$~fm$^3$ (mixed pairing prescription).  
The pairing parameter $t_0'$ is chosen so as to reproduce the empirical mean neutron gap 1.392 MeV of $^{120}$Sn extracted from the 5-point formula~\cite{Bender2000}. 
The angular momentum cut-off value is set to $19\hbar/2$.


At this point we should qualify our earlier statement about the role of the effective mass. 
We have seen that it hardly plays any role in bulk, static properties, like ground-state energies, of closed-shell nuclei~\cite{kids_nuclei1}.   
We will now examine also the two-neutron separation energy $S_{2n}$, which for a given isotope $(Z,N)$ is defined as  the difference 
\begin{equation} 
S_{2n}(Z,N) = E_B(Z,N) - E_B(Z,N-2) , 
\end{equation} 
where $E_B$ is the binding energy of the nucleus, a positive quantity for bound nuclei. 
Open-shell as well as closed-shell nuclei must be considered. 
We will use two different EoSs to demonstrate the possible effect of the effective mass assumptions.

\begin{figure}
\includegraphics[width=0.5\textwidth]{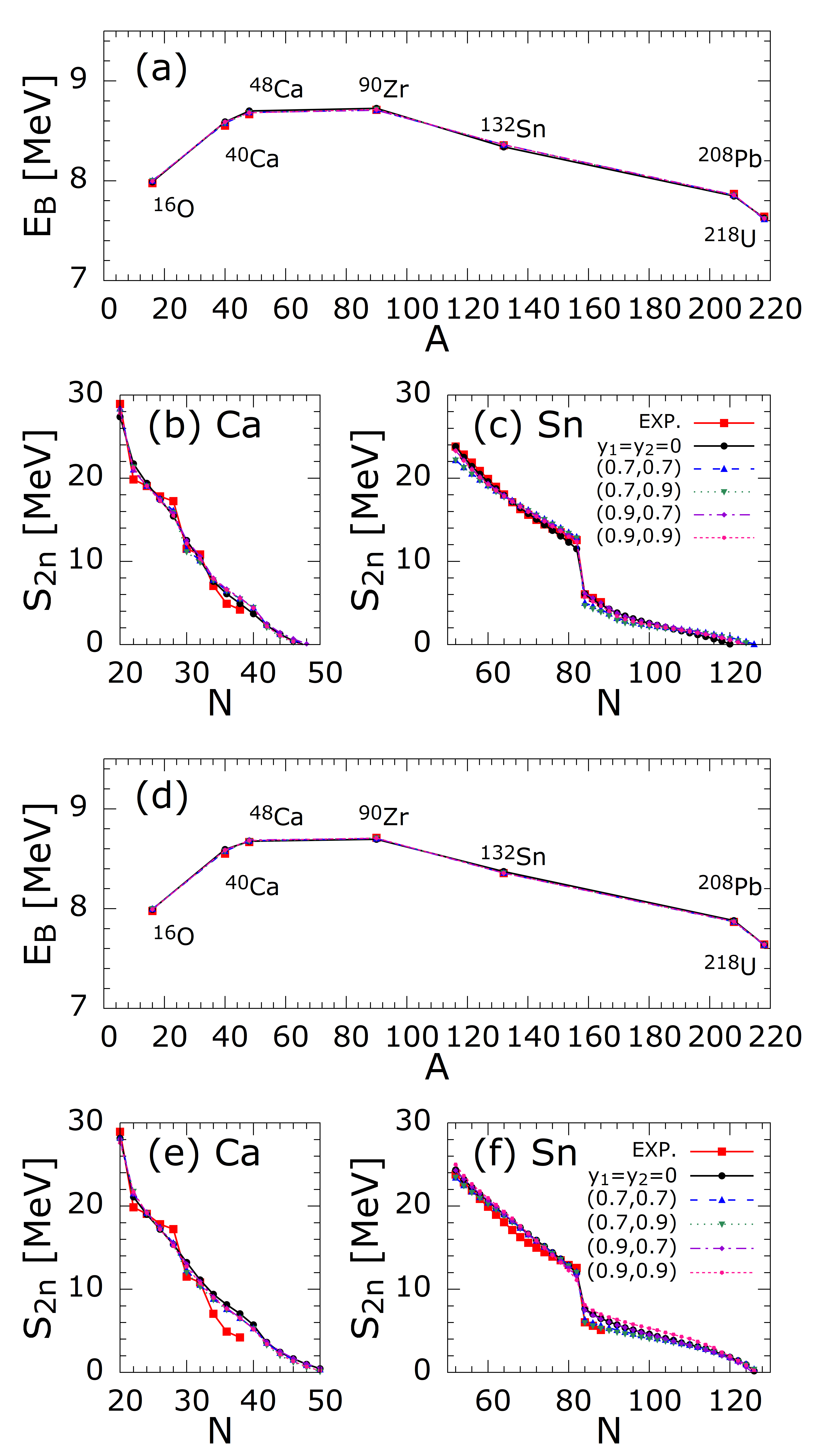} 
\caption{Dependence of results on kinetic terms (effective mass) illustrated for two EoS's. 
(a)$-$(c): KIDS-P4 EoS, $(J,L,K_{\mathrm{sym}}/K_{\tau},Q_{\mathrm{sym}})=(33,49,-157/-374,583)$~MeV (rounded values); 
(d)$-$(f): $(J,L,K_{\mathrm{sym}}/K_{\tau},Q_{\mathrm{sym}})=(30,50,-78/-300,650)$~MeV. 
(a), (d): Binding energy per particle $E_B/A$ of the indicated closed-shell nuclei; lines are drawn to guide the eye. 
(b), (e): Two-neutron separation energy $S_{2n}$ of Ca isotopes. 
(c), (f): $S_{2n}$ of Sn isotopes. 
The assumed effective mass values $(\mu_s,\mu_v)$ are indicated or the prescription $y_1=y_2=0$.  
 \label{Fig:mstar} }
\end{figure}

For two EoS parameterizations 
and for different prescriptions for $\mu_s,\mu_v$ we determine the  corresponding KIDS EDFs with the help of 13 input data (see Sec.~\ref{Sec:KIDS}) and a pairing parameter reproducing the $^{120}$Sn pairing gap. 
One EoS, labeled KIDS-P4 following Ref.~\cite{kids_nuclei2}, 
corresponds to $(J,L,K_{\mathrm{sym}}/K_{\tau},Q_{\mathrm{sym}})=(33,49,-157/-374,583)$~MeV (rounded values) and was fittd to the Akmal-Pandharipande-Ravenhall (APR) EoS~\cite{apr}.   
The second one corresponds to different symmetry energy parameters, 
$(J,L,K_{\tau},Q_{\mathrm{sym}})=(30,50,-77.635/-300,650)$~MeV.    
The possible effect of the effective mass assumptions  is illustrated . 
In Fig.~\ref{Fig:mstar}, we show results for the binding energy per particle $E_B/A$ of the input closed-shell nuclei and for $S_{2n}$ along the Ca and Sn isotopic chains by 1) using the simplified procedure with $y_1=y_2=0$ (resulting in high effective masses, $0.8-1.0$) and 2) assuming two different values of the isoscalar and isovector effective mass $(\mu_s,\mu_v) = (0.7,0.7),~(0.7,0.9),(0.9,0.7),(0.9,0.9)$.  
We compare with available data. 

As we can see, the $E_B/A$ of the closed-shell nuclei is practically unaffected by the choice of the effective mass.   
(The differences in the total $E_B$ for each nucleus and EoS are of the order of 1~MeV.) 
The $S_{2n}$, however, and the corresponding predictions for the position of the drip line can be somewhat affected by the choice of effective mass.  
The effect can be attributed to the pairing energy being non-monotonic between shell closures and influenced by the single-particle level density, in turn determined by (mainly) $\mu_s$.
Therefore, we will provide final results using $(\mu_s,\mu_v) = (0.7,0.7)$ and $(\mu_s,\mu_v) = (0.9,0.9)$ as representative values. 
Any spread in the final results will represent the uncertainty in our predictions.
 
\subsection{Underdetermined EoS's? \label{Sec:Correl}}

In Skyrme and relativistic mean-field models a certain interrelation among symmetry energy parameters has been revealed, namely an almost linear relation between $(3J-L)$ and $K_{\mathrm{sym}}$~\cite{Mondal2017}. 
The authors of \cite{Mondal2017} point out that the relation holds 
``for a class of interactions with quadratic momentum dependence and a power-law density dependence" 
and indeed it is easy to demonstrate the origin of the relation in the case of traditional Skyrme functionals which belong in the above class. 
We will show that the limited number of free EDF parameters enforces a strong relationship among the symmetry energy parameters. 

The Skyrme EoS for SNM can be written in shorthand as
\begin{equation} 
\EPA^{\mathrm{Sk}} (\rho,0) = f\rho^{2/3} + a_0\rho + a_{\gamma}\rho^{1+\gamma} + a_2\rho^{5/3}  \, . 
\end{equation} 
($f\approx 75$~MeV~fm$^2$ is shorthand for the usual kinetic factors.) 
In the above, we recognize four independent Skyrme-EoS parameters: $a_0$, $a_{\gamma}$ and $\gamma$ are directly related to $t_0$, $t_3$ and the exponent $\gamma$; $a_2$ is related to the isoscalar effective mass $\mu_s$~\cite{kids_nuclei1,dutra2012}, 
\[ 
a_2=\frac{3}{10}\left(\frac{3\pi^2}{2}\right)^{2/3}\frac{\hbar^2}{m\rho_0} (\mu_s^{-1}-1) 
\] 
or approximately $37.5~(\mu_s^{-1}-1)$~MeV~fm$^3/\rho_0$.  Thus in principle four EoS parameters could be independently adjusted. 
Typically, a constraint must be imposed on the effective mass value ($0.6 \lessapprox \mu_s\lessapprox 1$), so more precisely three EoS parameters can be adjusted. 
Those would be $\rho_0,\EPA_0, K_0$.   
(Indeed, a need to adjust $K_0$ established the use of a power law $\gamma <1$ historically~\cite{Blaizot1980}.) 

The Skyrme expression for the density dependence of the symmetry energy can be written as follows,  
\begin{equation} 
S^{\mathrm{Sk}}(\rho ) = g\rho^{2/3} + b_0\rho + b_\gamma\rho^{1+\gamma} + b_2\rho^{5/3}  \, , 
\end{equation} 
 i.e., it involves three additional parameters $b_i$, corresponding to $x_0, x_3$, and the isovector effective mass $\mu_v$. 
($g\approx 42$~MeV~fm$^2$ is shorthand for the usual kinetic factors.) 
So up to three EoS parameters (such as $J,L,K_{\mathrm{sym}}$) can be adjusted, again with the caveat that $\mu_v$ cannot assume arbitrary values -- so that only two of the parameters can be adjusted truly independently. In particular, 
\[ 
b_2=\frac{1}{6}\left(\frac{3\pi^2}{2}\right)^{2/3}\frac{\hbar^2}{m\rho_0} \left[ -3(\mu_v^{-1}-1) +4(\mu_s^{-1}-1) \right] 
\]  
or roughly $42~\left[- 3(\mu_v^{-1}-1) + 4(\mu_s^{-1}-1) \right]$~MeV~fm$^2/\rho_0$.  
In addition, given the strong contribution of the kinetic terms ($a_2,b_2$) to the density dependence, strong limitations must be considered for the effective mass to better control the EoS parameters and the neutron-star mass-radius relation~\cite{Tsang2019}.

The above analytical expressions and the definitions of $J,L,K_{\mathrm{sym}}$ lead to the following analytical relation between $K_{\mathrm{sym}}$ and $(3J-L)$,  
\begin{equation} 
K^{\mathrm{Sk}}_{\mathrm{sym}} = -3(1+\gamma )(3J-L) + (1+3\gamma ) g \rho_0^{2/3} + 2(2-3\gamma )b_2\rho_0^{5/3}, 
\label{Eq:KsymSk}
\end{equation} 
which is equivalent to Eq.~(13) of Ref.~\cite{Mondal2017}. There, a regression analysis involving numerous EoS parameterizations gave the result 
(rounded up here for simplicity)
\begin{equation} 
K_{\mathrm{sym}} \approx -5 (3J-L) + 67~\mathrm{MeV} \, .
\label{Eq:Mondal}    
\end{equation} 
If in Eq.~(\ref{Eq:KsymSk}) we set $\gamma \approx 1/3$ (keeping in mind that values between $1/6$ and $2/3$ are widely in use)  and $\rho_0\approx 0.16$~fm$^{-3}$  
and assuming, on average, 
$\mu_s\approx \mu_v \approx 0.8$ ($b_2\rho_0\approx 10.5$~MeV~fm$^2$), we get 
\begin{equation} 
K^{\mathrm{Sk}}_{\mathrm{sym}} \approx -4 (3J-L) + 32  ~\mathrm{MeV} \, ,
\label{Eq:MyMondal}
\end{equation}  
which does not deviate much from Eq.~(\ref{Eq:Mondal}). 
The two expressions are compared graphically in Fig.~\ref{Fig:Correl}(b). 

We conclude that even if ``nature prefers" to deviate from the above linear relation, the modeling may lack the flexibility to accommodate that preference.  
Then the Skyrme and similar models for the EoS are underdetermined. 
A similar conclusion was reached in Ref.~\cite{Mar2019} based on an analysis of about 50 models and a Taylor expansion of the symmetry energy around saturation density. 
With the KIDS model, which can take the form of an extended Skyrme functional with an adequate number of independent parameters, the relevant EoS parameters can be 
explored independently~\cite{kidsnm,kids_nuclei1,kids_nuclei2}. 
Such is the purpose of this work in relation to the symmetry energy and nuclear structure.  
We will return to the above relation in discussing our results, see Sec.~\ref{Sec:KJL}.

\section{Procedure: Successive steps of elimination and correction \label{Sec:Method}}

For the EoS of SNM we adopt a fixed, three-term KIDS parameterization~\cite{kidsnm} corresponding to a saturation density $\rho_0=0.16$~fm$^{-3}$, energy per particle at saturation $\EPA_0 = -16$~MeV, and nuclear incompressibility at saturation $K_0 = 240$~MeV. 
\tcb{We note that that there is still a sizable uncertainty of roughly $\pm 20$~MeV in the empirical knowledge of $K_0$~\cite{Garg2018}. 
Also discussed recently is a possible 1-2 MeV discrepancy in $\EPA_0$ with respect to ab initio calculations~\cite{Atkinson2020}. 
However, the focus of the present work is 
on the parameters characterizing the density dependence of the nuclear symmetry energy, and generally isovector quantities, 
for which the empirical uncertainties are larger. 
Consequently, a highly accurate description of SNM is of secondary importance at the present stage.} 
We focus on the post important symmetry-energy parameters, namely $J,L,K_{\mathrm{sym}}$. 
The skewness parameter of the symmetry energy will be kept fixed to a given value $Q_{\mathrm{sym}}=650$~MeV, because 
we have found that moderate variations of $Q_{\mathrm{sym}}$ (within 100 -- 200~MeV) have very little effect on nuclear structure and even a marginal effect on neutron-star properties~\cite{kids_nuclei2}. The above value for $Q_{\mathrm{sym}}$ comes from within the range  420 -- 750 MeV  determined from  fitting to 
microscopic equations of state denoted by APR \cite{apr} and QMC \cite{qmc} in \cite{kids_nuclei2}.

When we are examining nuclear structure, instead of $K_{\mathrm{sym}}$ we explore the corresponding liquid-droplet parameter $K_{\tau}$ 
given by~\cite{Roca2018} 
\begin{equation} 
K_{\tau} = K_{\mathrm{sym}} - \left(6+\frac{Q_0}{K_0}\right) L \, . 
\label{Eq:Ktau}
\end{equation} 
\tcb{For the current choice of SNM parameters this gives $K_{\tau} = K_{\mathrm{sym}} -4.446L$. 
Obviously, } 
switching between $K_{\tau}$ and $K_{\mathrm{sym}}$ is straightforward. 

With the SNM EoS parameters and $Q_{\rm sym}$ fixed, we proceed to explore the quantities of interest, 
$J$, $L$ and $K_{\rm sym}$ or $K_\tau$. 
For their values, we adopt ranges similar to or broader than those denoted by ``CSkP" in \cite{dutra2012}. 
In particular, we begin by using  $J=31,32,33,34$~MeV, $30~{\rm MeV}\leq L \leq 70$~MeV, and $-550$~MeV$\leq K_{\tau}\leq 0$~MeV or 
$K_{\rm sym}$  between $-400$ and $+300$~MeV. 
\tcb{The goal is not necessarily to shift the CSkP ranges, but to narrow them further, especially for $K_{\mathrm{sym}}$.}

\subsection{Finite nuclei}

\subsubsection{Broad search: Best-performing $K_{\tau}$ values in closed-shell nuclei \label{Sec:coarse}} 

As outlined in Sec.~\ref{Sec:KIDSfm}, for each point on the $(J,L,K_{\tau})$ hyperplane the corresponding KIDS functional parameters $(t_i, y_i, W_0)$ are obtained using as input the desired EoS parameters and 6 data points.
The predictive power of each parameterization is tested at this stage on additional closed shell nuclei.  
Specifically, we inspect the ADPD, Eq.~(\ref{Eq:ADPD}), 
where 
the $N=13$ observables are presently the binding energies and charge radii of $^{16}$O, $^{40}$Ca, $^{48}$Ca,  
$^{90}$Zr, $^{132}$Sn, and $^{208}$Pb as well as the binding enery of $^{218}$U. 
Data for the energies are taken from the NNDC~\cite{NNDC} and data for the radii from Ref.~\cite{Angeli2013}. 
The purpose of using closed-shell nuclei as a primary filter is because they provide a more clean ``EoS signal", as it were: 
there is minimal uncertainty to the EDF predictions from approximations involving pairing correlations (see also Sec.~\ref{Sec:OSN}) and, especially important for charge radii, 
neglected \tcb{beyond-HFB} correlations which are necessary for reproducing isotopic shifts between shell closures~\cite{Blaum2008,Angeli2009,fayans2017}. 
 
\begin{figure}
\includegraphics[width=0.5\textwidth]{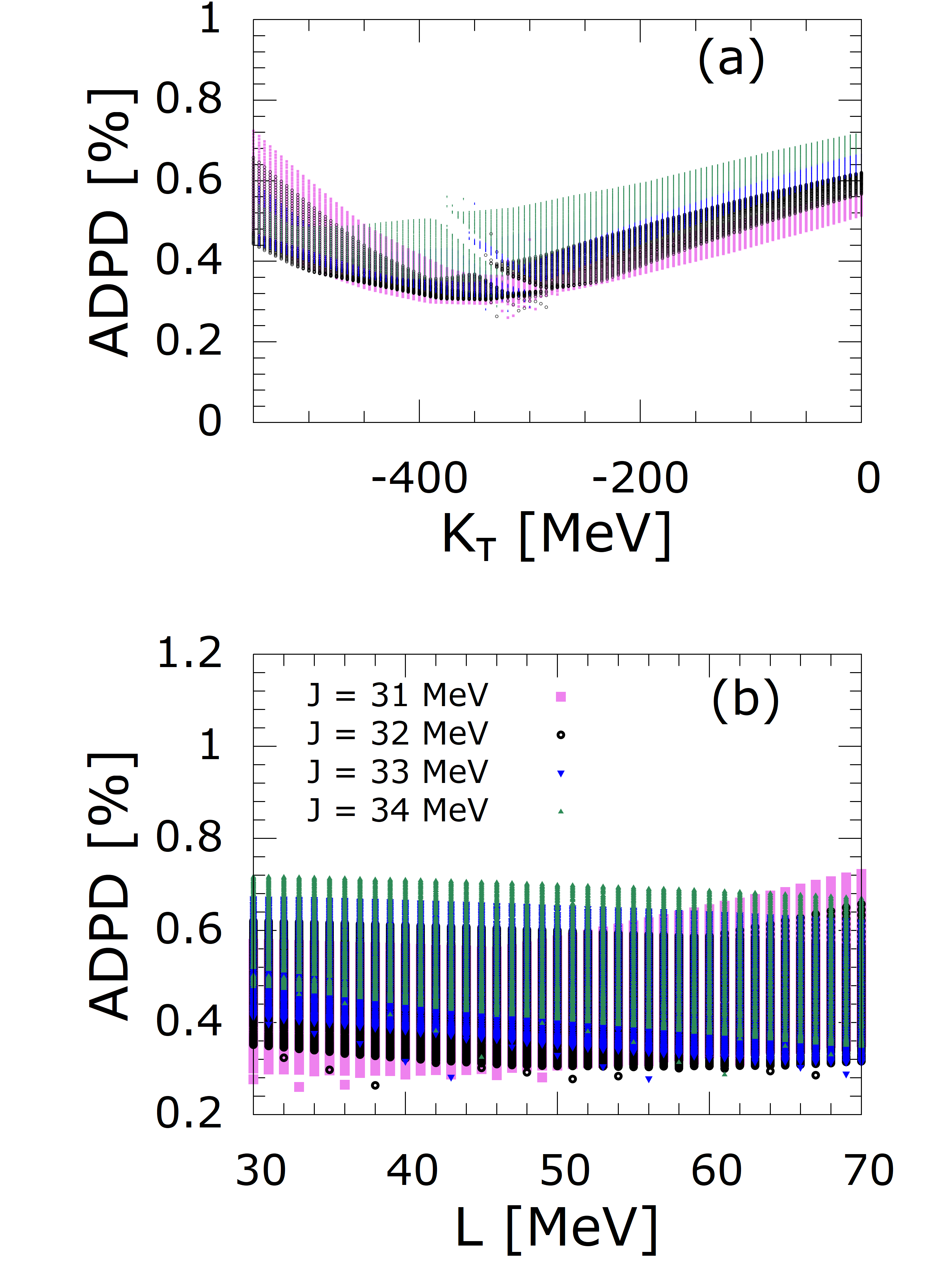} 
\caption{Average deviation per datum (ADPD) for the indicated $J$ values (a) as a function of $K_{\tau}$ and (b) as a function of $L$. 
The various points at given $(J,K_{\tau})$ correspond to different $L$ values, while the various points at given $(J,L)$ to different $K_{\tau}$ values. \label{Fig:ADPD} }
\end{figure}

Figure~\ref{Fig:ADPD} shows the ADPD for the indicated $J$ values as a function of $K_{\tau}$ (a) or as a function of $L$ (b). 
We observe that there is an optimal range of $K_{\tau}$ values roughly between $-400$ and $-300$~MeV (with some $J$ dependence). 
By comparison, the performance is quite flat with respect to $L$. 
Although the results shown in Fig.~\ref{Fig:ADPD} were obtained with the simple procedure ($y_1=y_2=0$), we have verified that the conclusions do not change when we perform fits with specific values of $(\mu_s,\mu_v)$.

Lower values of $J$ seem favorable, so in the following we will extend our search to lower values of $J$. 
We will examine three representative values of $K_{\tau}$ within the optimal range, namely $-400, -350, -300$~MeV.

%

\subsubsection{Refined search: Best-performing $L$ for given $(J,K_{\tau})$ in closed-shell nuclei \label{Sec:fine}} 
  
\begin{figure*}
\includegraphics[width=\textwidth]{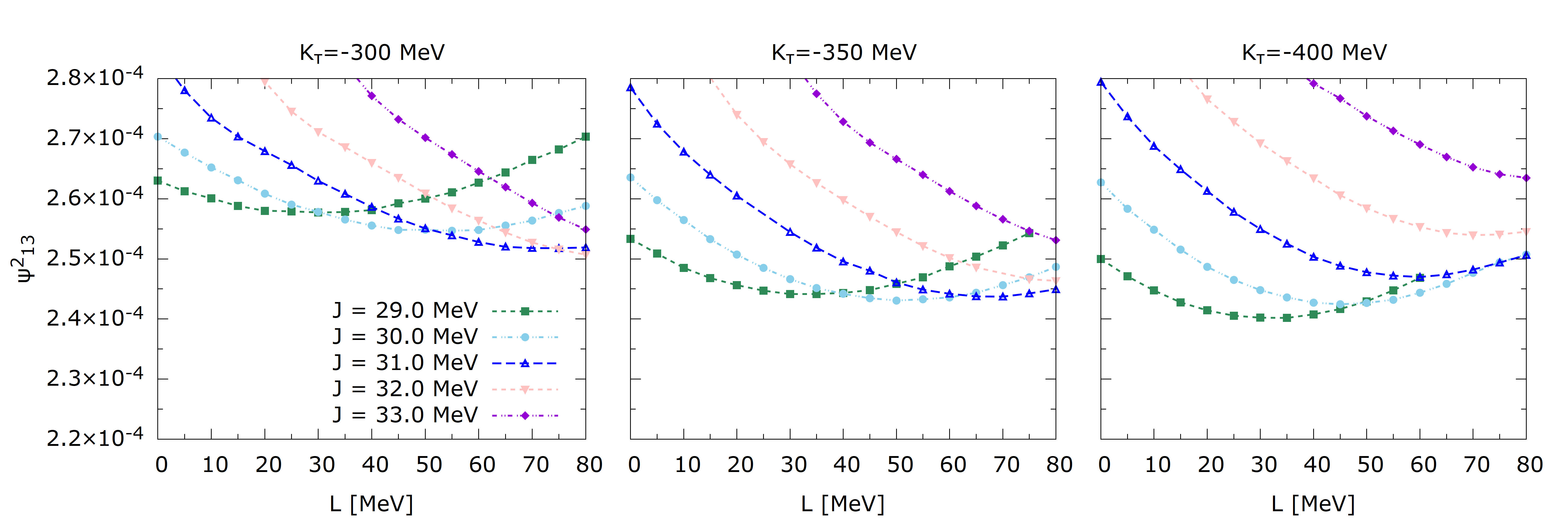} 
\caption{Fitting quality on 13 data as a function of $L$ considering the shown values of $J,K_{\tau}$ and $(\mu_s,\mu_v)=(0.9,0.9)$.  \label{Fig:chi2} }
\end{figure*}

For the representative values of $K_{\tau}=-400,-350,-300$~MeV we obtain KIDS-EDF parameterisations this time using all the 13 data mentioned above and in Sec.~\ref{Sec:KIDS}, i.e., the ${\psi^2}$ value is given by Eq.~(\ref{Eq:ChiSq}) with $N=13$. 
We consider $J=29,30,31,32,33$~MeV and $L$ up to 80~MeV in steps of 1 MeV. 
Results for the cost function ${\psi^2}$ are shown in Fig.~\ref{Fig:chi2} assuming $(\mu_s,\mu_v)=(0.9,0.9)$.  
The choice $(\mu_s,\mu_v)=(0.7,0.7)$ yields somewhat higher ${\psi^2}$ values, but similar trends. Based on the results for ${\psi^2}$ calculated on 13 data, we can determine the most-favored $L$ value for each $(J,K_{\tau})$ set. The resulting $(J,L,K_{\tau})_{\rm CS}$ sets, where ``CS" stands for closed-shell nuclei, are, in units of MeV: 
$(29,31,-300)$, $(30,55,-300)$, $(31,72,-300)$, 
$(29,33,-350)$, $(30,51,-350)$, $(31,68,-350)$, 
$(30,43,-400)$, $(31,57,-400)$, $(32,70,-400)$. 
We could of course include the $(29,33,-400)$ set which gives the lowest ${\psi^2}$. 
\tcb{However, we anticipated and we verified that the low values of $(J,L,K_{\mathrm{sym}})$  give very small neutron-star radii (well below 11~km for a 1.4$M_{\odot}-$mass star), so we discard the set in advance to keep the information load wieldy in the following.
} 

The choice of data inevitably introduces some bias in the above procedure. To control for the bias we will also consider variations of $L$ by 5~MeV. 
In other words, next we examine the performance not only of the above $(J,L,K_{\tau})_{\rm CS}$ sets, but also of the $(J,L\pm 5~{\rm MeV},K_{\tau})_{\rm CS}$ sets. 
\tcb{We note that our choice not to consider values of 80~MeV or higher will prove to be justified by the results of the next steps.}

\subsubsection{Bias check: Sn isotopic chain \label{Sec:Sn}} 

We now calculate the ADPD, Eq.~(\ref{Eq:ADPD}) using the $N=20$ measured binding  energies of Sn isotopes. 
We have checked that the performance of an EoS on the energies along any isotopic chain like Sn or Pb is representative of the performance along the other isotopic chains as well.    
(In particular, we have examined the O, Ca, Ni, Zr, Sn, and Pb chains.) 
We test all 27 EoSs with $(J,L,K_{\tau})=(J,L\pm 5~{\rm MeV},K_{\tau})_{\rm CS}$ and assuming $(\mu_s,\mu_v)=(0.7,0.7),~(0.9,0.9)$ -- that is, 54 EDFs in total. 
The results are tabulated in Table~\ref{Tab:Sn}. 
We use the notation $L_{\pm} = L_{\rm CS}\pm5$~MeV for any given  $(J,L,K_{\tau})_{\rm CS}$ set. 
\begin{table*}[h] 
\begin{tabular}{|r|c|c|c|}
\hline 
\hline 
$K_{\tau}$ [MeV]                           &  $-300$                                                      &          $-350$                           & $-400$                                   \\  
\hline\hline 
          &      $(J,L_-/L/L_+)$:(resp. ADPDs[$\%]$)                   &            $(J,L_-/L/L_+)$:(resp. ADPDs[$\%]$)          &            $(J,L_-/L/L_+)$:(resp. ADPDs[$\%]$)                                   \\ 

\hline 
  $(\mu_s,\mu_v)=(0.7,0.7)$                &  (29,26/31/36):(0.264/0.297/0.341)      &  (29,28/33/38):(0.368/0.426/0.485)  & (30,38/43/48):(0.462/0.525/0.590)  \\ 
                                 & (30,50/55/60):(0.362/0.423/0.484)       &  (30,46/51/56):(0.435/0.497/0.560)  & (31,52/57/62):(0.489/0.552/0.611)  \\  
                                                       & (31,67/72/77):(0.428/0.490/0.822)       & (31,63/68/73):(0.501/0.564/0.624)   &  (32,65/70/75):(0.505/0.565/1.085) \\ 
\hline \hline 
$(\mu_s,\mu_v)=(0.9,0.9)$     & (29,26/31/36):(0.159/0.216/0.278)      &  (29,28/33/38):(0.305/0.368/0.433)   & (30,38/43/48):(0.400/0.468/0.536)  \\ 
                               & (30,50/55/60):(0.291/0.358/0.426)       &  (30,46/51/56):(0.371/0.438/0.506)  & (31,52/57/62):(0.422/0.492/0.564)  \\  
                                                       & (31,67/72/77):(0.355/0.425/0.495)       & (31,63/68/73):(0.436/0.506/0.578)   &  (32,65/70/75):(0.434/0.506/0.580) \\ 
\hline 
\end{tabular}
\caption{ADPD results representing the performance of the $(J,L,K_{\tau})_{\rm CS}$ and the $(J,L_{\pm},K_{\tau})_{\rm CS}$ EoSs on the energies of Sn isotopes. 
Two blocks of results, each spanning three rows,  correspond to different effective masses as indicated.  
Three blocks of results, each spanning three columns, correspond to different $K_{\tau}$ as indicated. 
In each row of a $(\mu_s,\mu_v)\times K_{\tau}$ block we tabulate the values of a $(J,L_-/L/L_+)_{\rm CS}$ triad in MeV and the respective ADPD values.   
\label{Tab:Sn}}
\end{table*} 
Two trends are apparent: First, higher effective masses are favored by this set of data. Nevertheless, we will use both pairs of effective mass values in our final results, Sec.~\ref{Sec:Results}. 
Second, lower (higher) $L$ values are favored (disfavored). 
All $L_{\rm CS}+5$~MeV sets give large ADPDs compared to the other sets, with the exception of the $(29,36,-300)$~MeV set. 
Therefore, we next discard all but one of the EoSs with  $(J,L+5~{\rm MeV},K_{\tau})_{\rm CS}$ and henceforth  examine only the 
 $(J,L,K_{\tau})_{\rm CS}$,  $(J,L-5~{\rm MeV},K_{\tau})_{\rm CS}$ EoSs 
and in addition the  $(29,36,-300)$~MeV EoS. 
We  will denote the latter set of EoS's with the subscript ``CS+OS", where  OS stands for open-shell nuclei.

\subsection{Neutron star mass-radius relation and tidal deformability \label{Sec:NS}} 

The maximum mass of a neutron star has long been a key issue to discuss the stiffness of EoSs at high densities.
In the 1990's, a widely accepted upper limit of mass from observations was about $1.5 M_\odot$.
Masses larger than $1.5 M_\odot$ were reported, but the uncertainty was also large.
In the new millenium, many new observations have been conducted, and now the largest mass of pulsars is increased to more than $2 M_\odot$.
This large mass tends to favor stiff EoSs.
However, there are many models that produce a maximum mass larger than $2M_\odot$, 
so the large mass observation is not sufficient to constrain the EoS at high density.
Moreover, the baryon density at the center of a star with mass $2 M_\odot$ is more than $5 \rho_0$, 
so it is uncertain or unlikely that nucleons will be the only or the major constituents of the matter at that high density.
There are still many possibilities and large uncertainties in the EoS of neutron stars whose mass is larger than $1.4 M_\odot$.

Detection of the gravitational waves and evaluation of the tidal deformability with them introduced more powerful and
stringent constraints to the EoS of nuclear matter at high densities.
Masses of neutron stars in the merger of GW170817 were estimated around $1.4 M_\odot$~\cite{gw2017high}. 
Density at the center of $1.4M_\odot$ star is estimated at about $3 \rho_0$.
\tcb{It was shown in~\cite{Lim2015} that even if hyperons are created at densities below $3\rho_0$ and they soften the EoS, their effect to the mass of $1.4M_\odot$ star is negligible.
At the same time, additional exotic degrees of freedom, e.g. deconfined quarks, could appear around $3\rho_0$.
}
Therefore measurement of the $1.4 M_\odot$-mass star will be more critical and informative in obtaining a reliable EoS at densities 
below $3 \rho_0$.

\begin{figure}
\begin{center}
\includegraphics[width=0.5\textwidth]{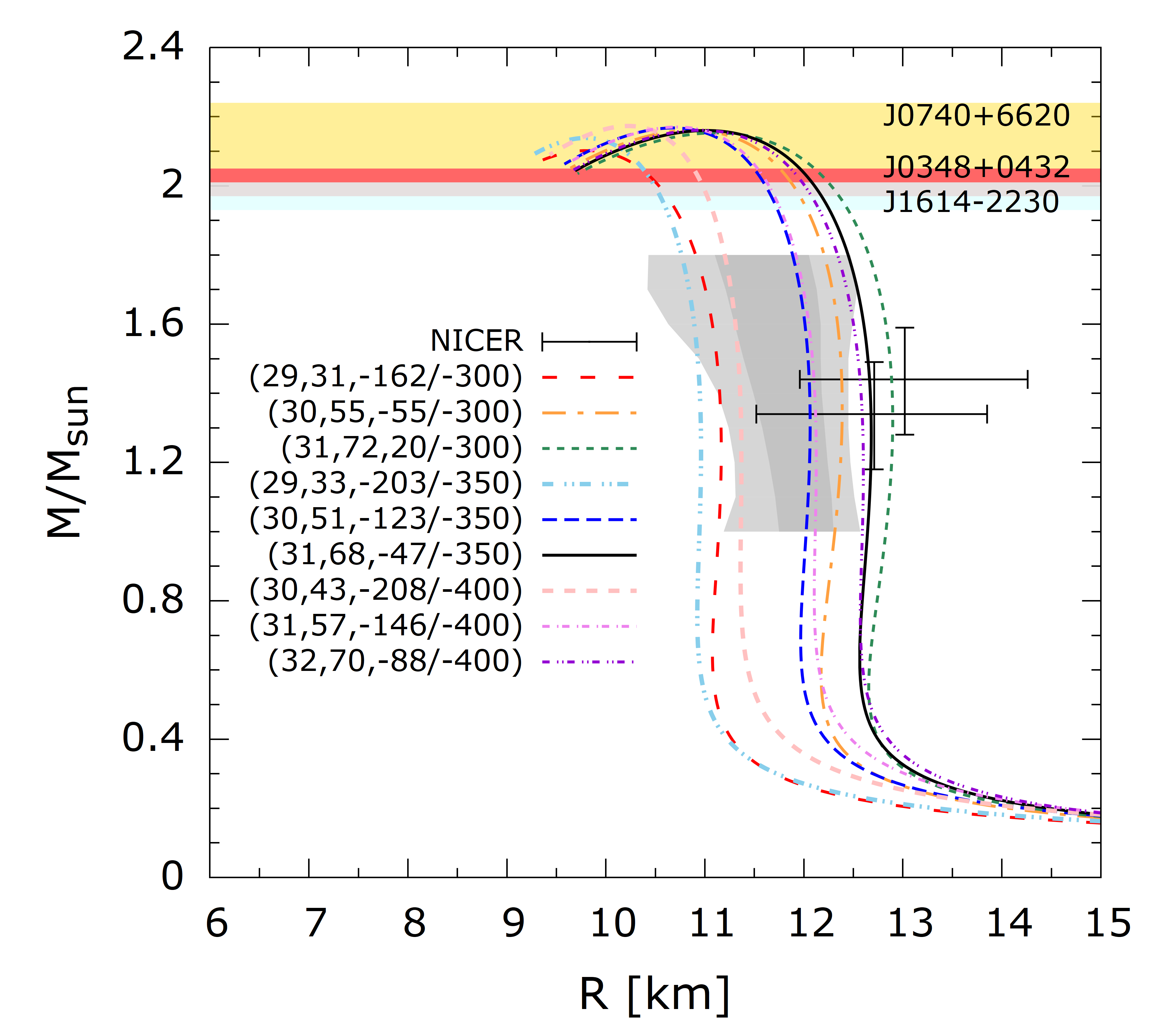}
\end{center}
\caption{Mass-radius relation with symmetry energy parameters $(J, L, K_{\rm sym}/K_\tau)_{\rm CS}$ noted in the panel in units of MeV.
Observational data include large masses in horizontal bands, X-ray burst analysis in gray band, and two bars with errors from NICER observation.}
\label{fig:nsmr}
\end{figure}
\tcb{ 
We proceed to examine the properties of neutron stars by using EoSs determined in this work. For simplicity, we use a crustal EoS based on the SLy4 functional. Of course, a consistent treatment of the crust EoS can affect the prediction for the radius of a neutron star~\cite{Ferreira2020}. However, our focus is on constraints for 1.4$M_\odot-$mass stars or heavier, for which the effect of the crust EoS on the radius (a couple of hundred meters) is too weak to be meaningful in the context of the present study given the existing constraints.
}
In Fig. \ref{fig:nsmr}, we plot the neutron star mass-radius relation with the nine previously selected sets of $(J, L, K_{\rm sym})_{\rm CS}$ values.
The bands around $2 M_\odot$ denote the range of large masses from observations \cite{mass1, mass2, mass3}.
Interestingly, the predicted maximum masses are distributed in a narrow range $(2.1-2.2) M_\odot$.
The irregular gray band represents the allowed mass-radius range obtained from an analysis of X-ray burst \cite{xburst},
and points with error bars are the result from the observation of a soft X-ray source in the NICER project \cite{nicer1, nicer2}.
In the result of NICER, two analyses are performed independently, and they show slightly different results.
Overlapping ranges of the two NICER results are $(1.3-1.5)M_\odot$ for mass and $(11.9-13.8)$ km for radius.
NICER gives radius relatively larger than those from the X-ray burst.

Looking at the results of theory, three EoSs corresponding to $(J, L, K_{\rm sym}) = (29, 31, -162),\, (29, 33, -203),$ and $(30, 43, -208)$~MeV are
inconsistent with the NICER result.
In addition, EoSs of $(29, 31, -162)$ and $(29, 33, -203)$~MeV can hardly satisfy the X-ray burst range around the mass $1.4 M_\odot$.
This may imply that an $L$ value smaller than 30 MeV can be ruled out by the constraints from both NICER and X-ray burst.
For $(30, 43, -208)$, it is within the range of X-ray burst, but does not enter the uncertainty range of NICER.
In the region of large radius, on the other hand, theory predictions are consistent with the range of NICER, but X-ray burst analysis 
rules out the $L$ values larger than 70 MeV.
This result is consistent with the upper limit of the range denoted by CSkP in \cite{dutra2012}, where $L({\rm CSkP}) = 48.6-67.1$ MeV.
Consequently, from the nine examined EoSs, the EoSs that satisfy both X-ray burst and the two NICER results simultaneously are reduced to
$(J, L, K_{\rm sym}) = (30, 55, -55),\, (30, 51, -123)$ and $(31, 57, -146)$~MeV.
The ranges that satisfy all conditions from neutron stars (consistent with NICER data 
and with X-ray burst data) and closed-shell nuclei 
can be tentatively assigned as 
\begin{equation}
(J, L)_{\rm CS+NS} \approx (30 - 31,50-65)~{\rm MeV}\, .
\label{eq:optJL}
\end{equation}
The range is consistent with the range of CSkP.
Among the three successful  EoSs, the $(30, 55, -55)$ set predicts a radius slightly larger than those of the other two EoSs 
for the $1.4M_\odot$ mass stars.
Tidal deformability is sensitive to the radius of a star, so it is worthwhile to consider what the theory predicts.

\begin{table*}[h]
\begin{center}
\begin{tabular}{l|c|c|c|c|c|c|c|c|c}\hline\hline
$K_\tau$  [MeV] & \multicolumn{3}{c|}{$-300$} & \multicolumn{3}{c|}{$-350$} & \multicolumn{3}{c}{$-400$} \\ \hline
$(J, L)$  [MeV] & (29, 31) & (30, 55) & (31, 72) & (29, 33) & (30, 51) & (31, 68) & (30, 43) & (31, 57) & (32, 70) \\ 
$K_{\rm sym}$  [MeV]  & $-162$ & $-55$ & $20$ & $-203$ & $-123$ & $-47$ & $-208$ & $-146$ & $-88$ \\ \hline
\hline
Mass [$M_\odot$] & 1.40 & 1.41 & 1.40 & 1.40 & 1.39 & 1.40 & 1.40 & 1.39 & 1.41 \\
$\Lambda$ & 234.7 & 455.0 & 604.9 & 209.2 & 402.3 & 535.3 & 256.9 & 401.7 & 485.8 \\
$R$ [km]  & 11.2 & 12.4 & 12.9 & 11.0 & 12.1 & 12.7 & 11.4 & 12.1 & 12.6 \\
$\rho_{\rm cen}$ [fm$^{-3}$] & 0.58 & 0.43 & 0.39 & 0.60 & 0.46 & 0.41 & 0.54 & 0.46 & 0.43 \\
\hline\hline
\end{tabular}
\end{center}
\caption{Tidal deformability $\Lambda$, radius $R$, and density at the center $\rho_{\rm cen}$
of the $1.4M_\odot$ neutron star with the $(J, L, K_{\rm sym}\, {\rm or} \, K_\tau)_{\rm CS+NS}$ values.}
\label{tab:td}
\end{table*}
Table \ref{tab:td} presents the properties of the star with mass close to $1.4 M_\odot$ in detail.
Considered quantities are the tidal deformability $\Lambda$, the radius $R$ in units of km and the density at the center $\rho_{\rm cen}$ in fm$^{-3}$.
Observational value of $\Lambda$ for the $1.4 M_\odot$ mass star from GW170817 was $\Lambda \leq 800$ in the first report \cite{gw2017high}.
An updated analysis provides a reduced range $190^{+390}_{-120}$ \cite{gw2017low}.
A new observation in 2019, GW190425 provides an upper limit $\Lambda \leq 600$ for the low-spin prior \cite{gw2019}.
All the symmetry energy parameters in Tab.~\ref{tab:td} satisfy the upper limit $\Lambda \approx 600$.
However, the magnitude of $\Lambda$ is clearly dependent on and correlated to the value of $L$.

For $(J, L, K_{\rm sym}) = (29, 31, -162), \, (29, 33, -203)$ and $(30, 43, -208)$~MeV, $\Lambda$ is roughly $200-250$.
In the opposite extreme $\Lambda$ is larger than 500 for $(31, 72, 20)$ and $(31, 68, -47)$~MeV.
For the three EoSs that satisfy both X-ray burst and two NICER results simultaneously, $\Lambda$  is roughly $400-450$.
Aside from the $(32, 70, -88)$ set, $\Lambda$ appears correlated to $L$ in a way that
$\Lambda \lesssim 300$ for $L \leq 40$, $\Lambda \approx 400$ for $L$ at the order of 50,
and $\Lambda \gtrsim 500$ for $L$ larger than 60.
Therefore, accurate measurement of the gravitational waves could play an essential role in reducing the uncertainty in the
density dependence of the symmetry energy.

The radius of the $1.4M_\odot$ mass star determined from GW170817 is $9.1-12.8$ km.
All the sets of symmetry energy in Tab.~\ref{tab:td} predict the radii within this range.
A softer EoS (or a small $L$) gives a larger value of $\rho_{\rm cen}$.
Three $(J, L, K_{\rm sym})$ sets that have $\Lambda$ at the order of 200 give the central density in the range $(3.4-3.7) \rho_0$.
Two stiff EoSs corresponding to $(31, 72, 20)$ and $(31, 68, -350)$ have the center densities $2.5 \rho_0$ and $2.6 \rho_0$, respectively.
For the three EoSs that are consistent with both X-ray burst and two NICER, theory predicts $\rho_{\rm cen} = (2.7-2.9)\rho_0$.
Similar to the tidal deformability, densities at the center are clearly classified according to the stiffness of the symmetry energy.

\begin{figure}[h]
\begin{center}
\includegraphics[width=0.5\textwidth]{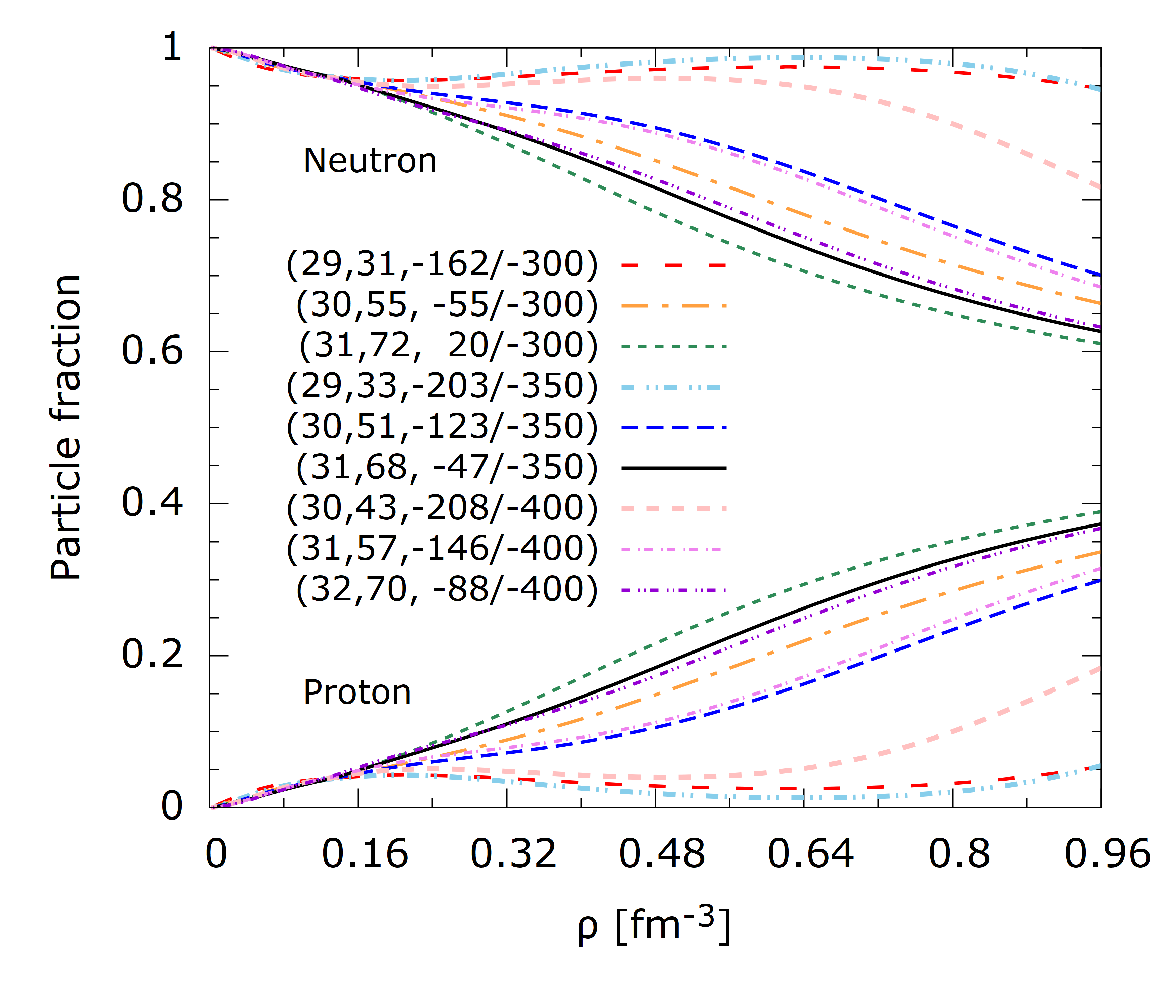}
\caption{Particle fraction in the core of neutron stars with the indicated $(J,\, L,\, K_{\rm sym}/K_\tau)_{\rm CS}$ values in MeV.}
\label{Fig:fraction}
\end{center}
\end{figure}

Figure \ref{Fig:fraction} shows the particle fraction in the core of neutron stars with the EoSs determined by 
$(J,\, L,\, K_\tau)$ in Fig. \ref{fig:nsmr}.
Fractions of neutrons and protons have a critical effect to the cooling of neutron stars.
Large fraction of the proton satisfies the energy-momentum conservation in the direct Urca process,
$n \rightarrow p e \bar{\nu}_e$ and $p e \rightarrow n \nu_e$, and it leads to a hyper-fast cool down of the neutron star.
In the simplest consideration, direct Urca is accessible when the proton fraction is larger than 1/9.
If direct Urca happens in the core, neutron star cooling calculated with any nuclear model cannot explain the
observation data \cite{cooling1, cooling2}.
Therefore, direct Urca should be prohibited at densities where nucleons are dominant constituents of the matter,
e.g. up to $(3-4) \rho_0$ which are center densities of the $1.4 M_\odot$ neutron stars.
Focusing on the densities 0.5 -- 0.6 fm$^{-3}$ in Fig. \ref{Fig:fraction}, one can easily categorize the proton fraction ($Y_p$) into
three classes: a class with high $Y_p$ values around 0.2, a middle class with $Y_p \approx 0.1$, and the last one with protons highly depleted.

Cooling is beyond the scope of this work, so we don't have explicit results of the cooling curve.
However, it is evident that the energy will be emitted via the direct Urca process in the neutron star of mass around $1.4_\odot$
for the high $Y_p$ EoSs $(J,\, L,\,K_{\rm sym})$ = (31, 72, 20), (31, 68, $-470$), and (32, 70, $-88$). 
If symmetry energy is stiff (i.e. large $L$), the cost to be asymmetric becomes expensive, 
so the matter favors more neutron-proton symmetric states.
For this reason, $Y_p$ is relatively large for the EoSs with large $L$ values.
Since $Y_p \geq 1/9$ is easily satisfied with large $L$ values, three large $L$ EoSs may not be 
allowed by the consistency with the cooling data of neutron stars. 
Thus we confirm again that the $L$ value is unlikely to reach 70~MeV or higher. 

So far we tested only the nine $(J,L,K_{\rm sym})_{\rm CS}$ sets on neutron stars and selected three of them, $(J,L,K_{\rm sym})_{\rm CS+NS}$. 
We have seen in Sec.~\ref{Sec:Sn} that the energies of open-shell nuclei favor lower values of $L$ with practically no deterioration in the description of closed-shell nuclei. 
Such a trend will certainly have effects to the properties of neutron stars, and acceptable ranges of $J$ and $L$ 
that are compatible with the astronomical observation. 
We will now examine the parameter sets $(J,L,K_{\rm sym})_{\rm CS+OS}$. 
%

\begin{figure}[h]
\begin{center}
\includegraphics[width=0.5\textwidth]{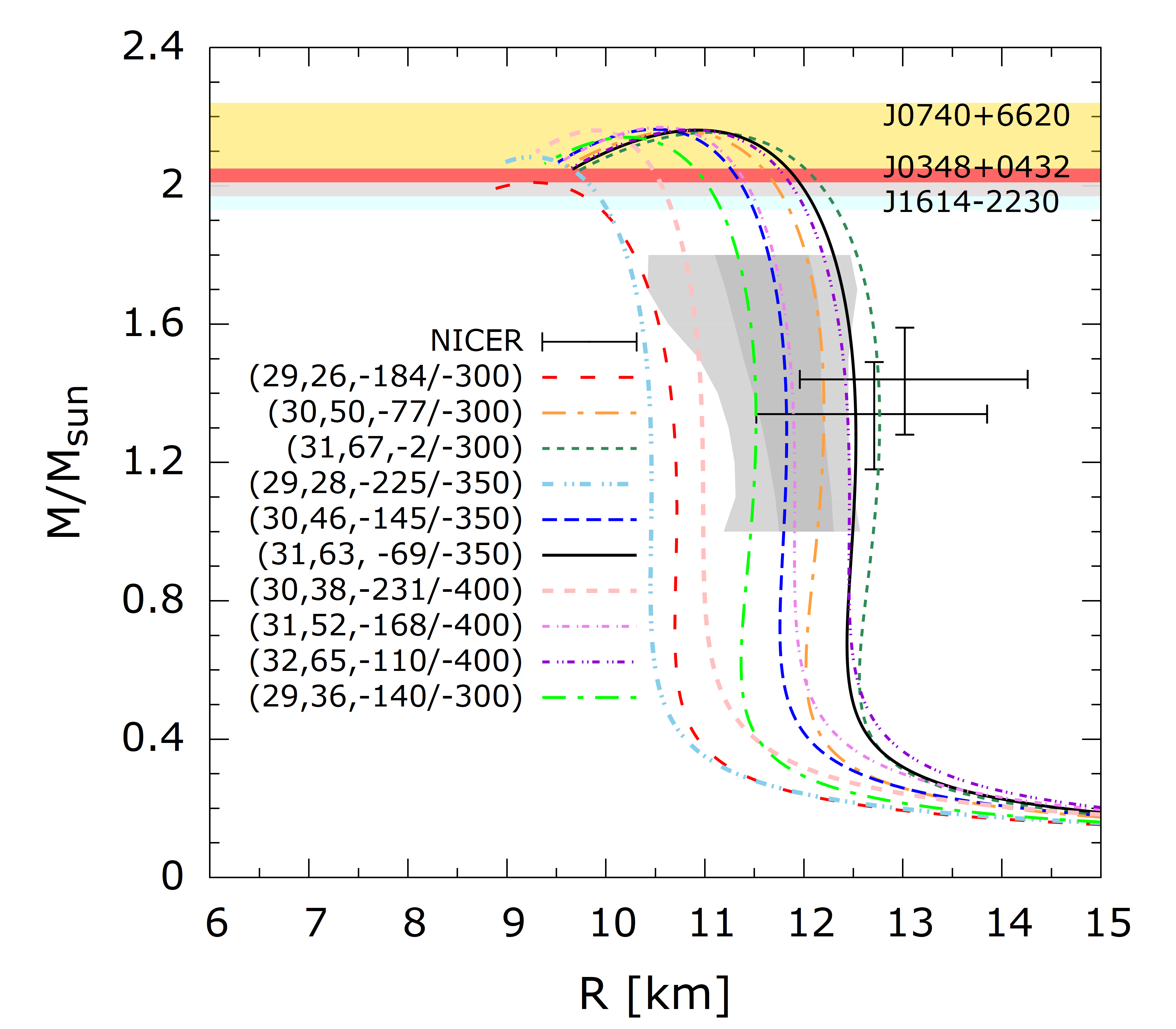}
\end{center}
\caption{Same as Fig.~\ref{fig:nsmr}, but for the ``CS+OS" parameter sets.}
\label{fig:nsmr2}
\end{figure}
Figure \ref{fig:nsmr2} shows the neutron star mass-radius relation for the ``CS+OS" sets. 
Compared to Fig.~\ref{fig:nsmr}, the curves for given $J,K_{\tau}$ are shifted to left as a whole. 
The three leftmost curves lie outside the acceptable ranges and therefore the corresponding EoSs can be discarded. 
The ($29,36,-300$) curve is marginally consistent only with the X-ray data and we discard it as well. 
We also discard the rightmost curve as inconsistent with the X-ray burst range. 
Given the current uncertainties, we tentatively accept the remaining five sets of $(J, L, K_{\rm sym})$ as reasonable. 
The surviving sets tentatively suggest 
the following ranges of parameters \tcb{so far within the KIDS framework}, 
\begin{equation}
(J,L)_{\rm CS+NS+OS} \approx (30-32,45-65)~{\rm MeV} \, ,
\label{eq:maxJL}
\end{equation}
slightly extending the range suggested in Eq.~(\ref{eq:optJL}). 
The new ranges are again consistent with the empirical range and CSkP of Ref.~\cite{dutra2012}. 
\tcb{That the $J$ and $L$ ranges are consistent with CSkP is understandable, because even traditional Skyrme functionals have sufficient free parameters to describe those two EoS parameters independently of each other.
It is also a welcome outcome that two different procedures (the survey of multiple functionals in Ref.~\cite{dutra2012} and the direct survey of the EoS parameters here) yield, in the end, consistent results} 

\begin{table*}[h]
\begin{center}
\begin{tabular}{l|c|c|c|c|c|c|c|c|c}\hline\hline
$K_\tau$ [MeV] & \multicolumn{3}{c|}{$-300$} & \multicolumn{3}{c|}{$-350$} & \multicolumn{3}{c}{$-400$} \\ \hline
$(J,\, L)$ [MeV] & (29, 26) & (30, 50) & (31, 67) & (29, 28) & (30, 46) & (31, 63) & (30, 38) & (31, 52) & (32, 65) \\ 
$K_{\rm sym}$ [MeV]  & $-184$ & $-77$ & $-2$ & $-225$ & $-145$ & $-69$ & $-231$ & $-168$ & $-110$ \\
\hline
Mass [$M_\odot$] & 1.41 & 1.41 & 1.40 & 1.41 & 1.40 & 1.40 & 1.39 & 1.41 & 1.40 \\
$\Lambda$ & 164.6 & 412.9 & 562.1 & 142.7 & 341.2 & 484.1 & 210.2 & 334.1 & 459.2 \\
$R$ [km]  & 10.7 & 12.2 & 12.8 & 10.4 & 11.8 & 12.5 & 11.0 & 11.9 & 12.4 \\
$\rho_{\rm cen}$ [fm$^{-3}$] & 0.66 & 0.45 & 0.41 & 0.68 & 0.49 & 0.43 & 0.59 & 0.49 & 0.44 \\
\hline\hline
\end{tabular}
\end{center}
\caption{The same as Tab.~\ref{tab:td}, but for the ``CS+OS" parameter sets..}
\label{tab:td2}
\end{table*}
Properties of $1.4M_\odot$ neutron stars for the $(J,L,K_{\rm sym})_{\rm CS+OS}$ sets are shown in Tab.~\ref{tab:td2}.
For the ranges of Eq.~(\ref{eq:optJL}), the tidal deformability of $1.4M_\odot$ neutron star has a range $400-460$.
With the maximal ranges Eq.~(\ref{eq:maxJL}), the acceptable range of the tidal deformability can be written as
\begin{equation}
\Lambda \approx 410 \pm 80.
\end{equation}
The radius and the central density are distributed over the ranges $(11.8-12.5)$ km, and $(0.43 - 0.49)\, {\rm fm}^{-3}$.
These ranges are slightly broader than, and include the ranges corresponding to the optimal ranges of Eq.~(\ref{eq:optJL}).

\section{Summary of constraints and results\label{Sec:Results}}

\subsection{Constraints on and correlations between symmetry-energy parameters \label{Sec:KJL}} 

\begin{figure} 
\includegraphics[width=0.50\textwidth]{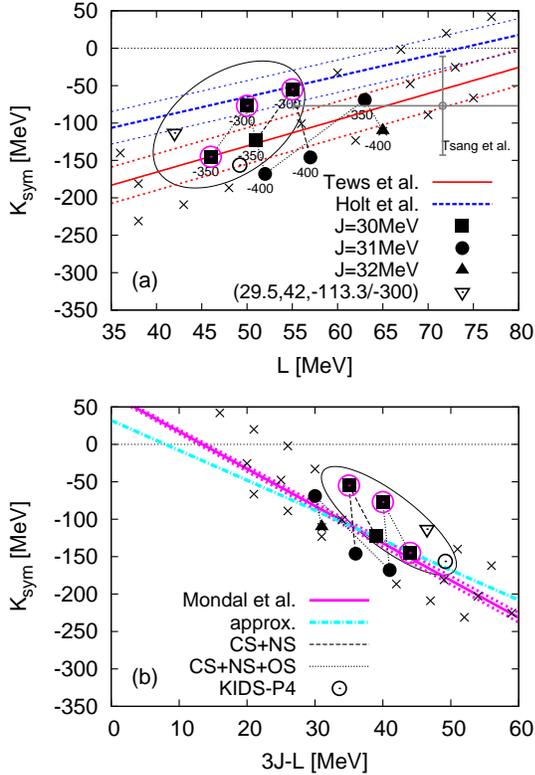}   
\caption{Constraints on and correlations between  (a) $K_{\mathrm{sym}}$ and $L$ and (b) $K_{\rm sym}$ and $(3J-L)$.  
The filled black points (squares, disks, triangles) correspond to our tentatively selected \tcb{KIDS} EoSs and they 
are connected according to the constraints they obey (CS: closed-shell nuclei; OS: open-shell nuclei; NS: neutron stars -- see text). 
In panel (a) the points are labeled also by the $K_{\tau}$ value in MeV. 
Filled points inside magenta circles correspond to the EoSs providing the best ADPD along the Sn chain, while 
crosses represent EoSs which were discarded in Sec.~\ref{Sec:Sn} and Sec.~\ref{Sec:NS}. 
The open circle corresponds to the KIDS-P4 EoS \cite{kids_nuclei2}.  
The inverted triangle corresponds to the newly examined point (see text) with the corresponding ($J,L,K_{\rm sym}/K_{\tau}$) values given in MeV.  
Ellipses circumscribe regimes free of rejected points.  
Lines depict correlations found in the literature (Tews et al.~\cite{Tews2017}, Holt et al.~\cite{lim2018}, Mondal et al.~\cite{Mondal2017}) except the line labeled ``approx." which corresponds to Eq.~(\ref{Eq:MyMondal}). 
The point with error bars labeled ``Tsang et al." represents the results reported in Ref.~\cite{Tsang2020}. 
 \label{Fig:Correl} }
\end{figure}

Let us now return to correlations such as the one  between $K_{\rm sym}$ and $(3J-L)$ discussed in Sec.~\ref{Sec:Correl}. 
That and other correlations reported in the literature~\cite{Mondal2017,Tews2017,lim2018,Li2020} are shown in Fig.~\ref{Fig:Correl} (lines)  along with our selected KIDS EoSs (filled black points). 
Uncertainty bands of the correlations are shown with dotted lines. 
The point with error bars labeled ``Tsang et al." represents the results reported in Ref.~\cite{Tsang2020} 
based on a Taylor expansion of the symmetry energy around the saturation density and neutron-star properties. 
The line labeled ``approx." corresponds to Eq.~(\ref{Eq:MyMondal}). 
The points representing the selected KIDS EoSs are connected by dotted lines according to the constraints they obey 
(CS: closed-shell nuclei; OS: open-shell nuclei; NS: neutron stars). 
Numbers in Fig.~\ref{Fig:Correl}(a) are the respective values of $K_{\tau}$ in MeV. 
Crosses represent  EoSs which were discarded in Sec.~\ref{Sec:Sn} and Sec.~\ref{Sec:NS}. 
Points inside magenta circles correspond to the EoSs providing the best ADPD along the Sn chain. 
The position of the KIDS-P4 EoS is also shown with a circle. 

The KIDS results shown in Fig.~\ref{Fig:Correl} strongly suggest that $K_{\rm sym}$ is negative and no lower than $-200$~MeV, in concordance with results based on chiral two- and three-body interactions~\cite{Drischler2016}. 
Ellipses have been drawn to circumscribe parameter  regions  free of discarded EoSs. 
The ellipses contain four of the best-performing EoSs explored in this work and they easily accommodate also the KIDS-P4 EoS, which is based on the APR EoS and has shown similar performance 
in properties of nuclei and neutron stars -- see Fig.~\ref{Fig:mstar}(a)-(c), Table \ref{Tab:Sn} and Ref.~\cite{kids_nuclei2}.  
$L$ values between about 40 and 55~MeV appear favored.  
Finally we remark that the sets obey the constraints from a lower bound on neutron-matter energy on the basis of unitary-gas
considerations~\cite{Tews2017}.

One should notice that, based on our results so far, it was not clear whether the ellipses should include the region around $(L,-K_{\rm sym})\approx (40-45,100-150)$~MeV or not, because we had no points therein (accepted nor discarded). 
Therefore we have examined one more point representative of that region, indicated with an inverted triangle in Fig.~\ref{Fig:Correl}. 
It turns out that the corresponding EoS, $(J,L,K_{\rm sym}/K_{\tau}) = (29.5,42,-113.3/-300)$~MeV, satisfies the neutron-star constraints, namely it is consistent with the X-ray burst and NICER data and the maximum mass. In addition, the ADPD values along the Sn isotopic chain are close to the best ones tabulated in Table~\ref{Tab:Sn}, 
namely $0.338\%$ and $0.269\%$ for $(\mu_s,\mu_v)=(0.7,0.7)$ and $(0.9,0.9)$, respectively. Therefore we accept that EoS as well into our final set. This brought the lower bound for our estimate of $L$ to about $40$~MeV, down from about 45~MeV. 

  
\tcb{The points representing the most likely values lie within the combined uncertainty bands shown in Fig.~\ref{Fig:Correl}(a). 
On the other hand, the correlation reported in Ref.~\cite{Mondal2017} and plotted in Fig.~\ref{Fig:Correl}(b) appears too restricting to constrain $K_{\rm sym}$ reliably, in the sense that most of the likely points (based on the present analysis) and especially the center of the ellipse are removed from the lines. 
We have already observed in Sec.~\ref{Sec:Correl} that the  dependence of $K_{\rm sym}$ on $J$ and $L$ in Skyrme models, Eq.~(\ref{Eq:KsymSk}),  enforces a model-specific correlation 
between $K_{\rm sym}$ and $3J-L$. 
We observe in Fig.~\ref{Fig:Correl}(b) that the analytical estimate of the correlation, Eqs.~(\ref{Eq:KsymSk}), (\ref{Eq:MyMondal}), labeled ``approx.", 
does not deviate much from the fitting result of Ref.~\cite{Mondal2017}. 
Our numerical results reinforce the suggestion that this correlation is model-specific.  
A more comprehensive correlation analysis based on the KIDS framework would be interesting in this respect. 
The ultimate goal is to determine the one true EoS as precisely as possible. 
} 

\subsection{Neutron drip line and neutron skin thickness \label{Sec:Drip}} 

We take the opportunity to report predictions for the position of the neutron drip line and for the neutron skin thickness of selected nuclei based on the 
 six 
EoSs contained in the ellipses in Fig.~\ref{Fig:Correl}.

We consider the drip nuclei focusing on even-even O, Ca, Ni, Zr, Sn, and Pb isotopes. 
For each isotopic chain the drip nucleus is defined as the last nucleus along the chain (highest neutron number,$N_d$) for which $S_{2n}$ is positive and the neutron Fermi energy is negative. 
In Table~\ref{Tab:Results} we summarize results obtained with the four selected sets of $(J,L,K_{\tau})$ values and with KIDS-P4. 
In all but five cases involving the heavier elements the result is independent of the effective mass choice. 
On the other hand, the drip isotope especially of the heavier elements may be influenced by the symmetry energy parameters. A dedicated study of the dependence would be of interest. 
%
\begin{table*}[h] 
\begin{tabular}{|r|cc|cc|cc|cc|cc|cc|}
\hline
     EoS     & \multicolumn{2}{c|}{ $(29.5,42,-300)$} &  \multicolumn{2}{c|}{ $(30,50,-300)$} & \multicolumn{2}{c|}{($30,55,-300$)} & 
\multicolumn{2}{c|}{($30,46,-350$)}    
 &  \multicolumn{2}{c|}{($30,51,-350$)}  
& \multicolumn{2}{c|}{KIDS-P4}    \\ 
     $(\mu_s,\mu_v)$        &  (0.7,0.7)  & (0.9,0.9)  &  (0.7,0.7) &  (0.9,0.9)  & (0.7,0.7) & (0.9,0.9)  & (0.7,0.7) &  (0.9,0.9)  &  (0.7,0.7) & (0.9,0.9)  &  (0.7,0.7) & (0.9,0.9) \\ 
\hline 
$N_d$: O       & 20   & 20    &       20   &  20                 &     20   &   20                  &     20  &  20                 &      20  & 20                   &    18   &  18                       \\    
            Ca      &  50   & 50   &      50  &   50                 &     54    &  54                  &     56   &   56                 &      58  &  58                  &    46   &  46                  \\   
            Ni       &  68   &  68  &       70   &  70                 &     70    &  70                 &     70    &  70                 &       72  & 72                  &     62   &   62              \\    
            Zr       &   94  &  94  &     96   &   96                 &     100  &   100               &   102   &  102                &      106 &  106               &     84   &   86                     \\   
            Sn      &   124 & 124   &     124   & 124                &     124   &  124               &    124   &  124               &      126 & 124               &     124    &  122         \\
            Pb      &  184   &  184  &    184   &  184                &     186  &  190               &    188   &  190               &      198 & 200               &      184   &  184                 \\   
\hline
 $r_{np}$: 
 $^{48}$Ca &  0.165  & 0.166   &    0.169 & 0.171                & 0.175  &  0.176           &   0.182 & 0.183              &  0.188 &  0.188         &   0.170       &  0.172                 \\    
 {[fm]}\hspace{1mm} 
 $^{68}$Ni   &  0.165  &  0.168  &   0.171 & 0.175               &  0.176 & 0.179            &   0.185  & 0.187             &   0.194  &  0.192        &  0.187         &  0.191           \\ 
 $^{132}$Sn &  0.214  & 0.219   &    0.224 & 0.229              &  0.231  &  0.237          &   0.242  &  0.247            &   0.250   &  0.255       &  0.242        &   0.249                  \\ 
 $^{208}$Pb &  0.151  &  0.153  &     0.160  &  0.163            &  0.167  &  0.169          &   0.176 &  0.178             &  0.183    &  0.185      &   0.180       &   0.183             \\   
\hline 
\end{tabular} 
\caption{For the KIDS functionals corresponding to the indicated EoS sets 
are shown 
for the drip line neutron number ($N_d$) of the shown isotopic chains and for the neutron skin thickness ($r_{np}$) of selected nuclei. 
The EoSs examined in this work (columns $2-6$) are labeled by the values of $(J,L,K_{\tau})$ in MeV. 
For all five, $Q_{\rm sym}=650$~MeV. 
KIDS-P4 has $(J,L,K_{\tau},Q_{\rm sym})=(33,49,-374,583)$~MeV. 
\label{Tab:Results}}
\end{table*} 

The O drip nucleus is generally predicted to have neutron number $N_d=18-20$ by many EDFs. 
Current data suggest that the last bound isotope is $^{24}$O, while $^{26}$O is only slightly unbound~\cite{Lunderberg2012}. 
On the other hand, it has been shown that the inclusion of realistic three-nucleon forces (as well as proper treatment of continuum states) can play an important role in describing particularly
the O drip nucleus~\cite{Hagen2012}. It is likely that Oxygen isotopes are too light to be described accurately by mean-field approaches. 

It is interesting also to comment on Ca. The discovery of $^{60}$Ca was published in Ref.~\cite{Tarasov2018}. As reported there, {\em ab initio} approaches wrongly predicted the drip line below $^{60}$Ca. EDF approaches which predicted the drip line closer to and beyond $^{70}$Ca agreed better with data. 
The present KIDS results also indicate that the drip nucleus is at $^{70}$Ca or beyond. 
Interestingly, the KIDS-P4 EDF, which was based on the APR EoS, represents an intermediate prediction for $N_d$ (as does, e.g., the SLy4 functional~\cite{Chabanat1998} with $(J,L,K_{\tau},Q_{\rm sym})\approx (32,46,-323,522)$ ). 
It is obvious from Fig.~\ref{Fig:mstar}(b),(e) that KIDS-P4 provides a better description of the Ca two-neutron separation energy than the $(J,L,K_{\tau})=(30,51,-350)$~MeV EDF. 
Precise fits of the EoS parameters along isotopic chains within the KIDS framework could be pursued in the future to provide more confident predictions. 

In Table~\ref{Tab:Results} the results for the neutron skin thickness are also reported for nuclei of current theoretical and experimental interest~\cite{Roca2018}. 
The variations among predictions of the various KIDS functionals are small considering the current experimental precision. 
Specifically, for given $J$ and $K_{\tau}$ the variation observed is less than 0.015~fm.  
This is not surprising given that the neutron skin thickness is understood to be correlated with $L$ and that the span in $L$ values considered here is less than 15~MeV. 
The results are consistent with various correlations reported in the literature, see, e.g., \cite{Roca2018} and references therein.
It is interesting to note the variations observed between the predictions with different $K_{\tau}$. 
The lower $K_{\tau}$ value ($-350$~MeV) consistently leads to thicker neutron skins. 
The trend becomes obvious when we consider the two EoSs in the third and sixth columns with equal $J$ and almost equal $L$ but different $K_{\tau}$.  
We conclude that the $K_{\tau}$ parameter is also important to consider in studies of the neutron skin thickness. 
A dedicated study would be worth pursuing in the future.

\section{Summary and perspectives\label{Sec:End}}

Using KIDS EoSs for unpolarized homogeneous nuclear matter at zero temperature and KIDS EDFs with pairing correlations in spherical symmetry 
we have explored the hyperplane of symmetry-energy parameters. 
Using both nuclear-structure data and astronomical observations as a testing ground, 
a narrow regime of well-performing parameters has been determined, free of non-physical correlations and constraints on the nucleon effective mass. 
Correlations reported in the literature between symmetry-energy parameters were critically discussed. 

\tcb{Our results based on the KIDS framework} strongly suggest that $K_{\rm sym}$ is negative and no lower that $-200$~MeV, that $K_{\tau}$ lies between $-400$ and $-300$~MeV 
and that $L$ lies between 40
 and 65~MeV, with $L\lessapprox 55$~MeV more likely. 
For $J$ we find most likely the values $31.0\pm 1.5$ MeV. 
\tcb{Our results represent a considerable narrowing of the fiducial range especially for $K_{\mathrm{sym}}$ with respect, e.g., to the CSkP range~\cite{dutra2012}.}  
For the selected well-performing sets, we reported corresponding predictions for the position of the neutron drip line and the neutron skin thickness of selected nuclei of current interest. 
The results are found only weakly affected by the choice of effective mass values. 
The results rather underscore the role of $K_{\mathrm{\tau}}$ and of precise astronomical observations. 

The somewhat heuristic step-by-step process followed here was necessitated by the diversity of the data and their uncertainties. 
There are several refinements which could be pursued in the future in order to pin down a smaller domain of realistic $(J,L,K_{\rm sym})$ values (the ultimate goal being a point, \tcb{as long as} one accepts that the nuclear EoS is unique). 
First, full advantage can be taken of nuclear data by fitting the EoS parameters to many measured masses along isotopic chains. 
\tcb{The sensitivity of the results on the treatment of pairing correlations should be examined more carefully in that case.}
Having already determined approximately a reasonably small regime of EoS parameters which describes well closed-shell nuclei and neutron stars, the task becomes feasible. 
In the process, we could also explore the role of $Q_{\rm sym}$. 
However, based on a previous study~\cite{kids_nuclei2}, it is not expected to affect our results considerably. 
More-precise astronomical observations, leading to better constraints for the high-density EoS, might be required for constraining the $Q_{\rm sym}$.    
The EoS of symmetric matter could also be explored.   
\tcb{A comprehensive statistical analysis of parameter correlations taking into account the uncertainties in our knowledge of the various observables can also be considered.}
In brief, more-precise constraints are possible with precise fits to nuclear energies and, in the future, more-precise input from astronomical observations.

\section*{Acknowledgments}
The work of HG was supported by the National Research Foundation of Korea (NRF)
grant funded by the Korea government No. 2018R1A5A1025563. 
YMK was supported by NRF grants funded by the Korea government (No. 2016R1A5A1013277 and No. 2019R1C1C1010571).
The work of PP was supported  by the Rare Isotope Science Project of the Institute for Basic Science funded 
by the Ministry of Science, ICT and Future Planning and the National Research Foundation (NRF) of Korea (2013M7A1A1075764).
CHH was supported by the National Research Foundation of Korea (NRF)
grant funded by the Korea government No. 2020R1F1A1052495.


%

\end{document}